# Putting GenAI on Notice: GenAI Exceptionalism and Contract Law

David Atkinson[1,2]

***Abstract***
Gathering enough data to create sufficiently useful training datasets for generative artificial intelligence requires scraping most public websites. The scraping is conducted using pieces of code (scraping bots) that make copies of website pages. Today, there are only a few ways for website owners to effectively block these bots from scraping content. One method, prohibiting scraping in the website terms of service, is loosely enforced because it is not always clear when the terms are enforceable. This paper aims to clear up the confusion by describing what scraping is, how entities do it, what makes website terms of service enforceable, and what claims of damages website owners may make as a result of being scraped. The novel argument of the paper is that when (i) a site's terms of service or terms of use prohibit scraping or using site content to train AI and (ii) a bot scrapes pages on the website including those terms, the bot's deployer has actual notice of the terms and those terms are therefore legally enforceable, meaning the site can claim a breach of contract. This paper also details the legal and substantive arguments favoring this position while cautioning that nonprofits with a primarily scientific research focus should be exempt from such strict enforcement.

---

[1] Assistant Professor of Instruction, Business, Government and Society Department, McCombs School of Business, University of Texas, Austin
[2] A special thanks to Saskia Reford, UT-Austin *juris doctor* candidate, for her invaluable contributions.

# Table of Contents

# I. Introduction

People are applying generative AI (GenAI), the kind of technology that can take a prompt from a user, like "write a poem about a cat tickling a dragon," and generate a related output, like "a girl was pulling a wagon / in the wagon a cat tickled a dragon…" to a wide range of uses, including coding, telemarketing, customer support, and education. Some consultancies, such as McKinsey & Company, claim GenAI could add $2.6 to $4.4 trillion to the global economy over the coming years.[3]

Several components undergird GenAI, including mountains of cash, gobs of computational resources (compute) needed to train GenAI models and allow them to perform tasks, and oceans of data. This last component, data, is the driving force behind this paper's analysis. All else being equal, data is the most important element of GenAI. If you must choose between the two, having high volumes of high-quality data is more important than having more compute, for example. Several factors determine what is high quality, including that the data are generally carefully crafted, incorporate varied and unique syntax/verbiage/expressions, others review the data to affirm it is good quality (*e.g.*, upvoted on Reddit, edited by professionals, or starred on GitHub), the data are longer and more complete (as opposed to short media posts), and they are in a standardized format. This is why GenAI training sets generally value content from the *New York Times* more highly than posts on *X*.

The most capable frontier models require copies of billions of web pages. GenAI companies convert the words, code, images, sounds, and videos from those web pages into trillions of tokens (sub-components of images and words). Those tokens are the training data used to train GenAI models. Meta's GenAI model, Llama 3.1, for example, trained on over 15 trillion tokens.[4] It stands to reason that similarly capable GenAI models, such as Gemini, ChatGPT, and Claude, require a similar volume of data.

---

[3]*The economic potential of generative AI: The next productivity frontier* (Jun 14, 2023), https://www.mckinsey.com/capabilities/mckinsey-digital/our-insights/the-economic-potential-of-generative-ai-the-next-productivity-frontier#key-insights

[4] *Introducing Llama 3.1: Our most capable models to date* (July 23, 2024), https://ai.meta.com/blog/meta-llama-3-1/. To provide some perspective, a million seconds is 11.5 days. A billion seconds is 31.7 *years*. A trillion seconds is 31,709 years. Fifteen trillion seconds would be 475,646 years.

The webpages with this data include most publicly available sites on the public internet, including YouTube, Reddit, GitHub, blogs, academic websites, news sites, and everything in between.[5] The GenAI entities and the companies they pay to work on their behalf use scraping bots to collect the data. These bots are pieces of software that perform tasks with little human oversight or involvement once deployed. The bots go from webpage to webpage and site to site, downloading the content on the web pages. In mere seconds, they can download all the content on websites, including massive sites with thousands of pages, making them far more efficient than a human performing the same task manually.

A common feature of websites, especially those with the highest quality data, is that they tend to have associated legal terms of service ("ToS" or "terms"). Often, these terms are only discoverable by clicking on a single hyperlink at the bottom of web pages, requiring users first to notice the terms and then to click on the link to view the terms. Courts commonly refer to this passive approach to ToS as browsewrap. Browsewrap contrasts with clickwrap, which requires affirmatively assenting to terms. The distinction and a more thorough discussion follow below. An additional wrinkle arises from licenses that generally accompany specific content on web pages, like images or code. These licenses are like mini ToS specific to copyright law, with their own requirements, restrictions, and permissions.

Legal matters are further complicated once we consider that contract law has almost exclusively focused on human or corporate actions. Yet, between 40% and 50% of all internet traffic is by bots.[6] Contract law was not designed for bots, but it may be time to significantly modify the law to acknowledge and adapt to their role.

The primary question this paper raises and answers is whether the standards of inquiry notice, as typically applied to humans visiting websites, should apply with equal force to bots who visit websites on behalf of humans. While the paper views this issue through a GenAI lens, the analysis and conclusion apply with equal force to any other scraping for any other purposes.

---

[5] Importantly, "publicly available" is not the same as "public domain". The former can still be protected by copyright laws. The latter is not.

[6] *See, e.g.*, *Bots Compose 42% of Overall Web Traffic; Nearly Two-Thirds Are Malicious, Reports Akamai* (Jun 25, 2024), PR NEWSWIRE, https://www.prnewswire.com/news-releases/bots-compose-42-of-overall-web-traffic-nearly-two-thirds-are-malicious-reports-akamai-302180377.html#:~:text=Bots%20Compose%2042%25%20of%20Overall,Thirds%20Are%20Malicious%2C%20Reports%20Akamai; Florian Zandt, *How Much Internet Traffic Is Generated by Bots?* (May 31, 2024), STATISTA, https://www.statista.com/chart/32339/share-of-web-traffic-caused-by-bots/ ("share of web traffic caused by bots" is 49.6%); *What is bot traffic? | How to stop bot traffic*, https://www.cloudflare.com/learning/bots/what-is-bot-traffic/#:~:text=It%20is%20believed%20that%20over,of%20that%20is%20malicious%20bots ("over 40% of all Internet traffic is comprised of bot traffic")

## II. How Entities Create GenAI

Describing the process of creating GenAI in detail is beyond the scope of this paper, but a helpful heuristic is to think of the process as a (simplified) supply train, as described by Grimmelmann and Cooper:

Step 1: Creation of expressive works or other information,

**Step 2: Collection and curation of enormous quantities of such data into training datasets (for GenAI models, these datasets are frequently scraped from the Internet by bots)**,

Step 3: Conversion of these expressive works or information into digitized data that can be interpreted by computers,

Step 4: Pre-training of a general, large-scale, base (also called foundation) generative-model architecture on these curated datasets,

Step 5: Fine-tuning the pre-trained base model on additional data to improve performance on a domain-specific task,

Step 6: Public release of the model's parameters, or embedding the model in a system for deployment in a software service, and

Step 7: End-user generation of outputs from a user-supplied prompt.

Step 8: Alignment of the model with human preferences or usage policies (a further stage of training that, for example, is responsible for ChatGPT behaving like a conversational chatbot)[7]

---

[7] Cooper, A. Feder and Grimmelmann, James and Grimmelmann, James, *The Files are in the Computer: On Copyright, Memorization, and Generative AI*, CHICAGO-KENT LAW REVIEW, (April 22, 2024). (lightly edited)

Readers do not need to understand the terminology in the preceding steps to understand the purpose of this paper. The emphasis on Step 2 above was added to identify the focus on where this paper fits into the GenAI development process. Steps 3-8 cannot occur without Step 2. Note also that because scraping occurs early in the supply chain, it is more likely that if it is illegal, everything that follows cannot be legal. Therefore, the question of bots and scraping is extremely important for GenAI as it currently exists. While it is difficult to parse out how much some company valuations, like Google's or Microsoft's, hinge on GenAI specifically because they provide an array of other products, we can be certain the value is not trivial. Some companies that only focus on GenAI, such as Cohere, Anthropic, xAI, and OpenAI, have valuations of $5.5 billion,[8] $40 billion,[9] $50 billion,[10] and $157 billion,[11] respectively. If the current approach to scraping is illegal, it could wipe out hundreds of billions of dollars' worth of valuation.

Notably, few people outside (or even inside) these GenAI companies know exactly what is in their datasets because the companies keep the training data secret. They may choose to do so for several reasons, including competitive advantage, protecting the privacy of individuals in the datasets, and potentially shielding the companies from claims as varied as copyright infringement and unjust enrichment. As this paper will demonstrate, the companies should also consider the role of breach-of-contract claims.

---

[8] Chris Metinko, *Cohere Raises $500M At $5.5B Valuation*, CRUNCHBASE NEWS (July 22, 2024), https://news.crunchbase.com/venture/ai-cohere-valuation-rises-psp-cisco-fijitsu/

[9] Kyle Wiggins, *Anthropic raises another $4B from Amazon, makes AWS its 'primary' training partner*, TECHCRUNCH (Nov 22, 2024), https://techcrunch.com/2024/11/22/anthropic-raises-an-additional-4b-from-amazon-makes-aws-its-primary-cloud-partner/

[10] Berber Jin, Tom Dotan, and Meghan Bobrowsky, *Elon Musk's xAI Startup Is Valued at $50 Billion in New Funding Round*, WSJ (Nov 20, 2024), https://www.wsj.com/tech/ai/elon-musks-startup-xai-valued-at-50-billion-in-new-funding-round-7e3669dc

[11] Cade Metz, *OpenAI Completes Deal That Values Company at $157 Billion*, N.Y. TIMES (Oct. 2, 2024) https://www.nytimes.com/2024/10/02/technology/openai-valuation-150-billion.html

## III. Enforceable Agreements

Broadly, two types of legal agreements on websites pertain to scraping: terms of service and licenses. These agreements differ from the contracts governing products offered by a website, typically governed by Master Service Agreements (MSA). There are a few primary types of ToS agreements: browsewrap, clickwrap, and sign-in-wrap. Importantly, while these descriptive distinctions of browsewrap, clickwrap, and sign-in-wrap are meaningful and used by courts, they are not the defining part of contractual analysis. In fact, courts have been known to mislabel agreements, referring to clickwrap agreements as browsewrap agreements, for example.[12] As the United States District Court for the Southern District of New York explains, these classifications "are notable, but not necessarily outcome determinative."[13] Ultimately, "the basic principles of contract law do not apply differently." This section will examine each type and then discuss licensing.

| Types of ToS Agreements | | |
|---|---|---|
| **Type of Agreement** | **Description** | **Screenshot Example** |
| Browsewrap | "Browsewrap simplifies the signing process even further by automatically assuming users have accepted an agreement if they use the website. Typically used for Terms of Service agreements, browsewrap agreements use notices like banners and hyperlinks to tell users about the contract."[14] | 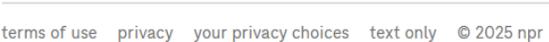 |
| Clickwrap | "Clickwrap agreements, also known as click-to-sign, click-accept, or clickthrough agreements, are online contracts that users can agree to by checking a box or clicking a button that indicates their consent."[15] | 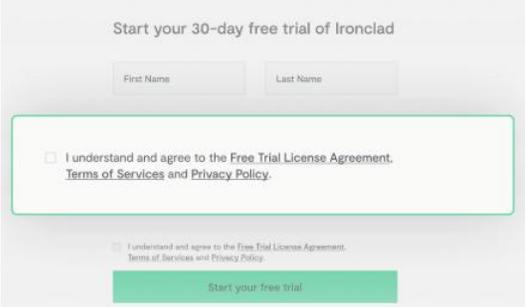 |

---

[12] Eric Goldman, *Ninth Circuit Enforces a Browswrap Agreement that was Actually a Clickthrough - Patrick v. Running Warehouse* (Feb. 16, 2024), https://blog.ericgoldman.org/archives/2024/02/ninth-circuit-enforces-a-browsewrap-that-was-actually-a-clickthrough-patrick-v-running-warehouse.htm.

[13] Plazza, 289 F. Supp. 3d at 548.

[14] *Clickwrap vs. Browsewrap: What's the Difference?*, IRONCLAD, https://ironcladapp.com/journal/contracts/clickwrap-vs-browsewrap/

[15] *Id.*

| Type | Description | |
|---|---|---|
| Sign-in Wrap | "Sign-in-wrap is a mix of clickwrap and browsewrap that collects contract acceptance through the performance of another action like signing up or logging in."[16] | 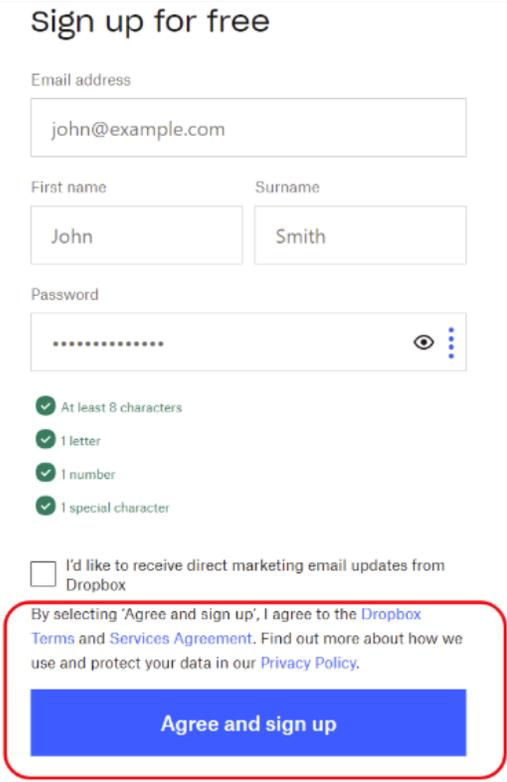 |
| License | "A Software License Agreement is a contract that allows a licensee to use software, but not own it. The software maker keeps some rights that the licensee doesn't get, like continuing to sell the software to others and granting the licensee permission to use the software on a lone computer.<br><br>The licensee also gets other rights and abilities, like the right to amend the code to make the software integrate better with other programs."[17] | 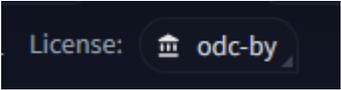 |

| Types of Notice | |
|---|---|
| **Type** | **Description** |

---

[16] *Id.*
[17] *What is a Software License Agreement?*, IRONCLAD, https://ironcladapp.com/journal/contracts/software-license-agreement/

| Notice | "Notice is the constitutional requirement that a party being brought into court be informed of the controversy."[18] |
|---|---|
| Inquiry Notice | "Notice that is imputed to a person at the time that existing information would prompt an ordinarily prudent person to investigate the issue further."[19] |
| Constructive Notice | Constructive notice is the legal fiction that someone actually received notice (being informed of terms that could affect their interest). If certain procedures have been followed, the law will consider a person to legally have received notice, even if in fact they did not. The notice under this doctrine arises by presumption of law from the existence of facts and circumstances.[20] |
| Actual Notice | "Actual notice is a notice that is given directly to a party or is personally received by a party informing them of terms that could affect their interests."[21] |

Terms of Service agreements are a creature of contract law and are thus governed by fundamental elements of contract formation. One of the most important elements is mutual assent, which refers to the idea that there is a "meeting of the minds" between the offeree and offeror of a contract.[22] The parties must agree to the same terms, conditions, and subject matters.[23] Mutual assent must be found for the ToS agreement to be enforceable, and it is often proved through the framework of offer and acceptance. One party offers certain terms, and the other party accepts; therefore, both parties know their continuing obligations to each other.

In website agreements, mutual assent can be harder to prove as there are fewer clear moments of "offer" and "acceptance." To accommodate the unique nature of the internet, Courts will infer mutual assent in certain situations with websites. If a user is on notice of the website's terms and continues to use the website, the user has functionally accepted the terms. If the terms were unacceptable to the user, the user could simply not use the website. By continuing to use the website after being on notice of the terms, the user has accepted the terms, which will be enforceable.

---

[18] Notice, LEGAL INFORMATION INSTITUTE, https://www.law.cornell.edu/wex/notice
[19] *Inquiry Notice,* QUIMBEE, https://www.quimbee.com/keyterms/inquiry-notice
[20] *See, Constructive notice*, LEGAL INFORMATION INSTITUTE, https://www.law.cornell.edu/wex/constructive_notice
[21] *Action notice*, LEGAL INFORMATION INSTITUTE, https://www.law.cornell.edu/wex/actual_notice
[22] Legal Information Institute, Mutual Assent (2023), https://www.law.cornell.edu/wex/mutual_assent#:~:text=Mutual%20assent%20refers%20to%20an,assent%20must%20be%20proven%20objectively.
[23] *Id.*

## A. *Browsewrap*

Browsewrap is by far the most common form of presenting terms of service on a website. This means that unless browsewrap is enforceable, most scraping of the internet by bots will not be subject to enforceable contract terms.

Browsewrap agreements do not require affirmative action on the user's part (the person visiting the website). The Ninth Circuit defines browsewrap agreements as a website offering terms through a hyperlink, and the "user supposedly manifests assent to those terms simply by continuing to use the website."[24] The offeree (the user) never consummates any action to affirmatively accept the terms given by the offeror (the website). Since the user takes no action, there is often no clear way to tell if the user was on notice of the terms, so courts are reluctant to enforce these agreements.[25]

However, it is not a bright-line rule that courts will *never* enforce browsewrap agreements. Courts can find these agreements binding if there is proof the user was or should have been on notice of the terms.[26] For browsewrap agreements, "actual notice" often cannot be found, which requires proof that the user was directly served notice of the terms (*e.g.,* by clicking on the hyperlink).[27] But, browsewrap agreements may satisfy "constructive notice," which is enough to enforce the terms. Constructive notice means that while the user did not directly receive notice of the terms (*i.e.,* "actual notice"), "the law will consider them to have received notice."[28] In the form of website agreements, constructive notice is satisfied through inquiry notice.[29] Inquiry notice consists of knowledge of facts and circumstances sufficiently pertinent to enable reasonably cautious and prudent persons to investigate. For ToS, inquiry notice is satisfied if the website puts a reasonable user on notice of the terms and conditions.[30] If a court finds that a reasonable user would have known about the terms, courts will assert that users constructively received notice. Because there is constructive notice, users functionally consented to the website's terms.

---

[24] Berman v. Freedom Fin. Network, LLC, 30 F.4th 849, 856 (9th Cir. 2022).

[25] Id.

[26] *See, e.g.,* Cairo, Inc. v. CrossMedia Servs., Inc., No. C 04-04825 JW, 2005 U.S. Dist. LEXIS 8450, at 4-5 (N.D. Cal. April 1, 2005); Register.com, Inc. v. Verio, Inc., 356 F.3d 393 (2d. Cir. 2004)

[27] Legal Information Institute, Actual Notice (2022), https://www.law.cornell.edu/wex/actual_notice.

[28] Id.

[29] Constructive notice can be found in many ways. The most common is through inquiry notice, and inquiry notice is what courts use for websites. However, there are other ways to satisfy constructive notice not discussed in this paper.

[30] Plazza v. Airbnb, Inc., 289 F. Supp. 3d 537, 548 (S.D.N.Y. 2018).

The analysis of whether there is inquiry notice, and thus constructive notice, depends on the design and content of the website.[31] If the hyperlink to the ToS is buried at the bottom of the page, in small font, or hidden with other links, courts will likely not enforce the terms.[32] There is no set design formula for what makes an agreement conspicuous. Instead, courts analyze conspicuousness on a case-by-case basis. However, as mentioned, most browsewrap agreements fail constructive notice as the terms and conditions are not conspicuous enough. An average user would not be put on notice; thus, courts cannot presume the user agreed to the terms.

The implications for scraping under this traditional analysis of browsewrap are immense. Suppose a GenAI entity deploys a bot that scrapes websites with content subject only to browsewrap. In that case, the terms are unenforceable, and the GenAI company could use the scraped information as they please, subject to non-contract claims that allow for affirmative defenses, like fair use.

## B. *Clickwrap*

Websites may use clickwrap agreements, which differ from browsewrap agreements by requiring affirmative action on the user's part. The user must physically click a button stating that they agree to the relevant terms and conditions.[33] This is usually done by clicking the "I agree" or checkbox button before proceeding.[34] By requiring action on the part of the user, these agreements are much more likely to be enforced than browsewrap agreements.

Though the user does not need to view the terms before clicking "I Accept," courts easily find inquiry notice because users must affirmatively agree to the terms, which means users necessarily learn that such terms exist. Moreover, courts find actual notice with clickwrap agreements because the website directly asks for and requires assent to the terms before the user is allowed to proceed on the website.[35]

---

[31] Nguyen v. Barnes & Noble Inc., 763 F.3d 1171, 1177 (9th Cir. 2014).

[32] *Id*.

[33] *See* Sergei Tokmakov, *The Evolution of Clickwrap, Browewrap, and Sign-in Wrap Agreements* (Aug. 26, 2023), https://terms.law/2023/08/26/the-evolution-of-clickwrap-browsewrap-and-sign-in-wrap-agreements/#Definition_and_Characteristics.

[34] A variation of click-wrap agreements is click-through agreements. These agreements usually have a "proceed" or "submit order" button with the fine print below stating that by proceeding, one agrees to the website's terms and conditions.

[35] Sergei Tokmakov, *The Evolution of Clickwrap, Browewrap, and Sign-in Wrap Agreements* (Aug. 26, 2023), https://terms.law/2023/08/26/the-evolution-of-clickwrap-browsewrap-and-sign-in-wrap-agreements/#Definition_and_Characteristics. (Arguing clickwrap agreements require explicit consent, which is actual notice); Nguyen, 763 F.3d at 1176.

## C. Sign-in Wrap

Finally, there can be sign-in-wrap agreements, a relatively recent designation in the legal sphere.[36] These agreements are a combination of browsewrap and clickwrap agreements. Unlike clickwrap agreements, users do not take affirmative action to explicitly agree to the terms and conditions by separately agreeing to them. However, the user may still assent to the terms when they sign up to use the website's services because when they create an account or log into the website, the website notifies them of the existence and applicability of the terms and conditions.[37]

Sign-in-wrap agreements are more likely to be enforced than browsewrap agreements but are less likely to be enforced than clickwrap agreements.[38] They are not presumptively enforceable.[39] To be enforceable, the references or links to terms must be conspicuous so users can be on alert that they are agreeing to terms and conditions.[40] For example, small, hard-to-read fonts will make these agreements unenforceable because users are not aware that signing up was consenting to terms and conditions.[41] As California Judge Do explains, the determination of enforceability "comes down to fine gradations of text, color, location, and other forms of context."[42] Moreover, the prompt (signing up to the service) must be spatially connected to the manifestation of assent to the terms and conditions.[43] If not, users will not be aware of the implications of signing up to the website. For example, the United States District Court for the Northern District of Illinois did not validate a sign-in-wrap agreement in one case because the registration and the hyperlink page of the terms were on different screens, and the hyperlink was not conspicuous.[44]

---

[36] Berkson v. Gogo LLC, 97 F. Supp. 3d 359, 399 (E.D.N.Y. 2015).

[37] *Id.*

[38] *Id.* at 400. (Explaining that sign-in wrap agreements have elements of clickwrap and browsewrap agreements, and courts are split on enforcement.) *See also* TopstepTrader, LLC v. OneUp Trader, LLC, No. 17 C 4412, 2018 U.S. Dist. LEXIS 64815, 2018 WL 1859040, at *3 (N.D. Ill. Apr. 18, 2018) ("Sign-in-wrap agreements never have the user take an affirmative action to explicitly agree to the terms of the site, but it does require some form of affirmative action by requiring the user to sign up for an account.")

[39] Eric Goldman, *California Appellate Court Rejects Poorly Executed "Sign-in Wrap" - Sellers v. Just Answers* (Feb. 13, 2022)*,* https://blog.ericgoldman.org/archives/2022/02/california-appellate-court-rejects-poorly-executed-sign-in-wrap-sellers-v-justanswer-guest-blog-post.htm.

[40] Eirc Goldman, *Tubi's ToS Formation Fails - Campos v. Tubi* (Mar. 12, 2024), https://blog.ericgoldman.org/archives/2024/03/tubis-tos-formation-fails-campos-v-tubi.htm.

[41] *Id.*

[42] Eric Goldman, *California Appellate Court Rejects Poorly Executed "Sign-in Wrap" - Sellers v. Just Answers* (Feb. 13, 2022)*,* https://blog.ericgoldman.org/archives/2022/02/california-appellate-court-rejects-poorly-executed-sign-in-wrap-sellers-v-justanswer-guest-blog-post.htm.

[43] *Id.*

[44] *Id.*

## D. Licenses

Beyond the varying terms of service agreements, websites can also use licenses. In general, a license is "a permission, accorded by a competent authority, conferring the right to do some act which without such authorization would be illegal."[45] A software license, for instance, is a copyright license that tells other people what they can and cannot do with your code.[46] If a license is not specified for a code file, general copyright law applies, meaning the creator retains all rights to the source code "and no one may reproduce, distribute, or create derivatives of [the] work."[47] Licenses are often treated separately from the rest of the code. For example, on Hugging Face and GitHub the software license is one of many visual pieces of information with the associated code, data, or model, presented as either a hyperlink in a list of hyperlinks or a small icon among other icons.

Because contract law governs licenses, the descriptions of how the courts analyze the terms described above apply.[48] Thus, some licenses will be enforceable, and others will not, depending on whether the licenses can satisfy aspects of contract formation, most notably mutual assent. Importantly, most licenses resemble browsewrap agreements. The software or image has a small hyperlink alerting one to the license, and websites assume that by using the content, one will obey the license. However, as mentioned above, browsewrap agreements are rarely enforced because of a lack of proof that the user consented to the terms.

## IV. Overview Of Scraping

Until GenAI rose to prominence, relatively few websites seemed to care about being scraped. In fact, most site owners likely did not even know they were being scraped. To the extent they did, they probably only knew of scraping bots as they related to Google search. Google's Googlebot crawls websites to create its search index, which makes those sites discoverable via Google's search engine. Because Google search is by far the most used search engine in the world, if a site chose to block Googlebot, they essentially made themselves invisible on the web. So, it is unsurprising that most sites *wanted* to be scraped by Googlebot.

---

[45] Black's Law Dictionary, Licenses (2024), https://thelawdictionary.org/license/.
[46] Github, Licensing a Repository (2024), https://docs.github.com/en/repositories/managing-your-repositorys-settings-and-features/customizing-your-repository/licensing-a-repository.
[47] Id.
[48] If courts treat licenses differently from typical contracts, that would give an unfair advantage to using a license. Instead of being treated as an unenforceable browsewrap agreement, these licenses would be their own beasts that courts enforce even absent notice of any kind. Courts may argue that this distinction is fair since licenses are often used for software and images. Since users on these sites are more familiar with those domains, they may be more likely to read the licenses regardless of whether they are conspicuous, so the licenses should be enforced. Or, courts may argue that licenses resemble copyright law more than contract law, which does not require mutual assent.

Another common bot is deployed by the nonprofit Common Crawl, which scrapes a few web pages from a large sample of websites on the public part of the public internet every month and makes those datasets freely available. The target audience was initially researchers, but the datasets have since become a common source of training data for AI models.

GenAI complicates the justifications for scraping by entities such as Common Crawl. Whereas in 2017 the director of Common Crawl touted that "from the layperson's perspective, it is not at all trivial at present to extract a specific website's content (that is, text) from a Common Crawl dataset. This task generally requires one to know how to install and run a Hadoop cluster, among other things. This is not structured data."[49] What the director could not have known was that modules like ChatGPT would emerge and thrive on unstructured data—the exact type that Common Crawl provides. Because Common Crawl does not limit access or use to research of nonprofit purposes, it has become a common source of training data for for-profit entities.[50] Still, despite Common Crawl scraping about three billion pages a month, GenAI companies want more data. This is why they create their own bots to supplement sources like Common Crawl.

Other uses of scraping bots include gathering pricing information for price comparison sites like Kayak, Expedia, and Google Flights and gathering contact information such as phone numbers, email addresses, names, titles, and more to create mailing lists or, more nefariously, to send spam, phishing emails, and robocalls.

Each bot has a unique name. So, for Google, there is Googlebot.[51] There is also Bingbot (Microsoft makes AI CoPilot), Applebot, Bytespider (by ByteDance, the Chinese company that owns TikTok), FacebookBot (Meta makes the Llama models), CCbot (for Common Crawl), Amazonbot, GPTbot and OAI-SearchBot (for OpenAI, which makes ChatGPT), and so on. Companies often have more than one bot, making it challenging for websites to know who is scraping them and for what purpose. For example, Anthropic has Claudebot, Claude-Web, and Anthropic-AI.[52] Also, most GenAI companies have their own bot(s), including small companies most people have never heard of.

---

[49] Kalev Leetaru, *Common Crawl And Unlocking Web Archives For Research*, FORBES (Sep 28, 2017), https://www.forbes.com/sites/kalevleetaru/2017/09/28/common-crawl-and-unlocking-web-archives-for-research/

[50] *See* Shayne Longpre et al., *The Responsible Foundation Model Development Cheatsheet: A Review of Tools & Resources* (2024) at 19 ("In the text domain, web scrapes from common crawl (commoncrawl.org), or OSCAR (https://oscarproject.org/) are the base ingredient for most pretraining corpora.") (internal citation removed), https://arxiv.org/abs/2406.16746.

[51] There are actually several Googlebots, but we'll keep it singular for simplicity.

[52]Jason Koebler, *Websites are Blocking the Wrong AI Scrapers (Because AI Companies Keep Making New Ones)*, 404 MEDIA (Jul 29, 2024), https://www.404media.co/websites-are-blocking-the-wrong-ai-scrapers-because-ai-companies-keep-making-new-ones/

While most scraping is limited to what is publicly available (*i.e.*, accessible on the public internet without an account or payment), some scraper bots can use JavaScript to fill out website forms (like creating an account) and scrape the gated content behind the form. Some companies, like Bright Data, which Meta and Nvidia have used, advertise that it can get around "blocks and CAPTCHAs" with its "Unlocker" product.[53] Additionally, some programs and application programming interfaces (APIs) enable bots to interact with sites as though the programs and APIs were merely operating a web browser the way a human might, which can cause the websites to believe a human is interacting with them and not merely autonomous software so the website blocks the activity less aggressively.[54]

There are some acknowledged best practices to scrape more ethically. Characteristics of ethical scraping include that the companies do not try to hide the bot's name or source; they don't overload site servers by sending too many requests too quickly; they don't take any additional steps like filling out forms, creating accounts, answering CAPTCHAs, or subverting paywalls; and they respect robots.txt.

## A. *Robots.txt*

There is one widely-known tool to block these bots: robots.txt. Robots.txt is a protocol developed in 1994 that allows website owners to request specific bots not access some or all of their web pages. Most GenAI companies claim their bots respect robots.txt files, meaning that when a scraping bot visits a website, it first checks the robots.txt file to see which pages, if any, the bot is disallowed from scraping, and then it scrapes pages that are not identified. If the website does not have a robots.txt file, the GenAI companies presume this generally means the site is not opposed to having all accessible content scraped. This can start to seem coercive for search indexing because sites may be left with two choices that do not feel like true choices at all: either allow the site to be indexed for search and scraped to train GenAI, or be invisible on the internet.[55] There is usually no way to limit a bot to one use contractually but not another while maintaining a publicly accessible website.

---

[53] https://brightdata.com/

[54] *What is content scraping? | Web scraping*, https://www.cloudflare.com/learning/bots/what-is-content-scraping/

[55] Google finally created Google-Extended to let sites allow indexing for search but not to train AI in September 2023. Most bots do not differentiate. OpenAI only started allowing sites to block its bot at all in August 2023 (https://www.theverge.com/2023/8/7/23823046/openai-data-scrape-block-ai). As of November 2024, Cloudflare's bot blocker did not apply to Google bots due to its dual use as a search engine, essentially giving Google a free pass to scrape for being complicated.

A notable feature (or bug) of robots.txt is that it places the onus of maintaining robots.txt on each website. While large companies may have a team to manage and monitor their robots.txt file, it is unlikely that most websites do or that the site owners are even aware of what robots.txt is and how to use it. Moreover, each site has the burden of disallowing each new scraping bot as it pops up. With robots.txt, if content providers do not know the bot exists, they can't block it without disallowing all bots. When Anthropic shifts from one bot to another, the sites are supposed to somehow be aware of the change, or else they are presumed to allow the new bot to scrape the site even when the old bot from the same entity was disallowed. This all contributes to why haphazard disallowance is common. Nearly twice as many sites block GPTBot as CCBot and nearly six times as many block GPTBot as FacebookBot.[56] For the preceding reasons, robots.txt is usually less restrictive than website terms of service.[57]

Importantly, following robots.txt is only voluntary, and it is not clearly illegal to ignore it. The robots.txt file cannot enforce itself. In fact, bad actors may prioritize the pages they are told not to scrape because, presumably, that is where the more interesting content may be. Moreover, because data is the most important element for highly-capable GenAI and the financial rewards for having a top-performing GenAI model can be so great, companies have strong incentives to ignore robots.txt and adopt the position that silence is consent.[58]

---

[56] Alex Bocharov et al., *Declare your AIndependence: block AI bots, scrapers and crawlers with a single click* (Jul 3, 2024), https://blog.cloudflare.com/declaring-your-aindependence-block-ai-bots-scrapers-and-crawlers-with-a-single-click/

[57] Shayne Longpre et al., *Consent in Crisis: The Rapid Decline of the AI Data Commons*, https://www.dataprovenance.org/Consent_in_Crisis.pdf at 8.

[58] *See, e.g.*, Cade Metz et al., *How Tech Giants Cut Corners to Harvest Data for A.I.,* N.Y. TIMES (Apr 8, 2024), https://www.nytimes.com/2024/04/06/technology/tech-giants-harvest-data-artificial-intelligence.html

To give one example of a prominent GenAI company ignoring robots.txt, look no further than Perplexity AI. Perplexity AI is valued at $3 billion, and that valuation is based on its ability to provide GenAI-generated responses to user queries in a style similar to a search engine, which Perplexity AI calls an answer engine.[59] To provide those answers, PerplexityBot scrapes content on relevant sites, feeds it to other GenAI models, and shares the outputs. Unfortunately for Perplexity AI, they were repeatedly caught ignoring robots.txt.[60]

Finally, robots.txt is a leaky solution. Robots.txt only disallows direct scraping, and this invites a not-terribly-inventive workaround. Suppose a site disallows scraping by Googlebot or GPTBot, but not all bots. If Common Crawl's CCbot (or some other entity that routinely scrapes the web) is not disallowed and then they publish the scraped data, Google or OpenAI can simply collect the data from Common Crawl's website. It is an easy, free workaround that allows companies to claim they are technically complying with the letter of everyone's robots.txt files (though certainly not the spirit).

## B. *Blocking All Bots*

Some sites have decided to skip constantly configuring robots.txt and instead block all AI scraping bots all the time. One company, Cloudflare, has made this process relatively easy by adding a simple toggle button to their products (including their free tiers) to block "AI Scrapers and Crawlers." This is notable for a few reasons, including the fact that about a fifth of all internet traffic runs through Cloudflare. They had already allowed customers to block AI bots even when they complied with robots.txt, and over 85% of its customers turned the feature on, suggesting that perhaps people had *never* been ok with being scraped for AI but simply didn't have an effective way or requisite knowledge to block it.[61]

---

[59] Elizabeth Lopatto, *Perplexity's grand theft AI*, THE VERGE (Jun 27m, 2024), https://www.theverge.com/2024/6/27/24187405/perplexity-ai-twitter-lie-plagiarism .

[60] *See* Dhruv Mehrotra and Tim Marchman, *Perplexity is a Bullshit Machine*, WIRED (Jun 19, 2024), https://www.wired.com/story/perplexity-is-a-bullshit-machine/; Tim Marchman, *Perplexity Plagiarized our Story About How Perplexity is a Bullshit Machine*, WIRED (Jun 21, 2024), https://www.wired.com/story/perplexity-plagiarized-our-story-about-how-perplexity-is-a-bullshit-machine/; Katie Paul, *Exclusive: Multiple AI companies bypassing web standard to scrape publisher sites, licensing firm says*, REUTERS (Jun 21, 2024), https://www.reuters.com/technology/artificial-intelligence/multiple-ai-companies-bypassing-web-standard-scrape-publisher-sites-licensing-2024-06-21/; Kali Hayes, *OpenAI and Anthropic Ignore Rule that Prevents Bots from Scraping Online Content*, BUSINESS INSIDER (Jun 21, 2024), https://www.businessinsider.com/openai-anthropic-ai-ignore-rule-scraping-web-contect-robotstxt

[61] Alex Bocharov et al., *Declare your AIndependence: block AI bots, scrapers and crawlers with a single click* (Jul 3, 2024), https://blog.cloudflare.com/declaring-your-aindependence-block-ai-bots-scrapers-and-crawlers-with-a-single-click/

While Cloudflare claims its AI model is highly accurate when identifying AI bots, even when AI companies try to conceal their purpose, some prominent companies still try to subvert these technical guardrails. As *404 Media* reported, Nvidia engineers discussed how to get around technical blockers when scraping websites that were trying to limit or block bot activity so Nvidia could scrape the data and train their own GenAI model.[62] Given the highly competitive atmosphere in GenAI for lucrative financial returns, it seems more likely than not that other GenAI companies, including those claiming to respect robots.txt, take similar actions.

Notably, if a site blocks all bots through a tool like Cloudflare but does not redundantly block all bots in robots.txt, the GenAI entities could technically still claim they are acting ethically by complying with robots.txt even as they subvert the higher-level, more encompassing technical protections. Therefore, a pledge to respect robots.txt may not be sufficient and should probably be viewed skeptically.

## V. Bots Are Not Agents

When Person A agrees to represent Person B or Entity B, Person A is called an agent, and Person B/Entity B is called the principal. The agent acts on behalf of the principal to help achieve the principal's goals. So, when you go to a restaurant, the waiters, dishwashers, and cooks are all agents of the restaurant.

Society has agency law in business because it is impossible for one person or a single entity to perform all the actions necessary for a business to flourish without people acting on its behalf. This relationship is also why the principal can be legally liable for actions taken by the agent when performing their duties of obedience, reasonable care, and loyalty.

Typically, an agent-principal relationship requires consent on behalf of both parties. Bots, as you can imagine, are not conscious or self-aware. Therefore, they cannot consent to anything. If they cannot consent, then they cannot be agents. If they are not agents, they must be viewed as mere tools under the controlling entity's direct control in the law's eyes because the bots only do what the deployers program them to do. This is true even though bots tirelessly execute the goals encoded into them once deployed without much oversight.

---

[62] Samantha Cole, *Leaked Documents Show Nvidia Scraping 'A Human Lifetime' of Videos Per Day to Train AI*, 404 MEDIA (Aug 5, 2024), https://www.404media.co/nvidia-ai-scraping-foundational-model-cosmos-project/

This means all actions a bot takes are no different than if the deployer of the bot (the entity that controls the bot's capabilities) had taken the actions themselves, similar to how a construction worker is responsible for whatever their hammer hits when they wield the hammer. There is no "Oops, my bot crashed your website, not me" defense. This, in turn, means that all laws that might apply to a human manually scraping a website apply equally to the human for all actions taken by a bot scraping on the human's behalf. And because the humans this paper is concerned with are most likely acting within their official duties on behalf of a GenAI company, the GenAI company can be vicariously liable for all actions of the bots. To arrive at any other conclusion is to believe in GenAI exceptionalism and to grant bots more freedom and less responsibility than we expect of the humans who intentionally deploy them. It would also mean that we are aware of loopholes in the law, and we are shifting the burden of respecting the wishes of content creators from the people intentionally taking the content to the people who are creating it, essentially blaming the victim.

Finally, even if bots were legally recognized agents, their activity would *still* be binding on the bot deployer as the principal. Comments like those of Rich Skrenta, the head of the prolific scraper Common Crawl, that "[R]obots are people too" would seem to lean into the argument that anything a bot does that would be legally binding on a human should also apply to the bots.[63]

## VI. Legal And Policy Discussion

This portion explores the practical implications of applying contract law to bots, including the role of robots.txt, what it would mean if a bot deployer intentionally avoids terms, and how actual notice applies to bot deployers. The final two sections touch on the intersection of copyright law—GenAI's most litigated area of law–and contract law, including the importance of expression and how copyright law may preempt contract law.

### A. *Robots.txt*

Relying on robots.txt alone when determining whether content on a website was intended to be used freely comes with dubious presumptions. For example, while GenAI companies treat the lack of a robots.txt file on a website or a bot not being disallowed as the equivalent of the website owner not caring if the site's content is scraped, this presumption may not be accurate in many or most cases. It is unclear why presumption in favor of consent should be the default or why the burden should be on the content creator or owner to opt out of being scraped versus having to opt in.

---

[63] Matt Levin, *The Economy and Ethics of AI Training Data*, MARKETPLACE (Jan 31, 2024), https://www.marketplace.org/2024/01/31/the-economy-and-ethics-of-ai-training-data/

There are obvious alternative and equally valid presumptions a company could make, including that the site may not have a robots.txt file because they are unaware of the protocol, or they may not have updated the robots.txt file recently because the thought has not occurred to them, or they intend to update it but it is a lower priority than other tasks, or they don't know what their robots.txt file is for, or that they can signal to bots not to scrape them, or they don't know that there are new bots to disallow, or the site owner does not know that bots are scraping them.

In the past, being scraped may not have seemed like a problem so long as the site's servers were not overloaded. Often, scraping was beneficial because it helped keep their content discoverable on the internet. However, the calculus may have changed with the advent of chatbots. Whereas before, a scraper would generally either benefit a site (by making it discoverable) or at least not cause noticeable harm, there are reasons to believe GenAI scraping can cause active harm to the websites the bots scrape (discussed below). The effects of a bot scraping a site fundamentally differ from those of a human visiting a site.

Additionally, when a robots.txt file blocks a bot (or all bots), it is reasonable to assume the site also has legal terms against scraping. The onus should be on the bot deployer to review the site's terms if it does not follow robots.txt. Robots.txt may be one of the most efficient ways for sites to indicate to bots that its terms likely prohibit scraping, and scrapers assume the risk of breach of contract when they ignore robots.txt.

### B. *Intentional Avoidance*

When a human visits a webpage, it is nearly impossible to know if they are aware of browsewrap terms but are simply ignoring them. This is not a problem with bots. Because bots will do what they are programmed to do, we can be certain that if a bot is programmed to scrape every page of a website, it will scrape every page of that website. But what if a bot is programmed to scrape every page *except* the ToS?

Intentionally avoiding legal terms should be construed as constructive notice. A bot deployer would have to know there are ToS to avoid them. Knowing there are terms provides the requisite inquiry notice to make the terms binding. Moreover, we must discard arguments based on bots not understanding the terms. Humans need not understand the terms for the terms to be binding.[64] They must merely understand that terms apply. Most people bound to terms online probably do not know what they mean or their implications (*i.e*., indemnification, limitation of liability, assignment, etc.). Even lawyers can struggle to understand ToS and how the various components of an agreement interrelate.

---

[64] *See* Toth v. Everly Well, Inc., No. 23-1727 (1st Cir. Sept. 25, 2024)

Again, to argue that because GenAI bots do not understand terms, and therefore the terms should not be enforceable even when the bots are performing actions that would certainly make the terms enforceable against humans, would be to argue for GenAI exceptionalism specifically and bot exceptionalism more generally.

### C. *Actual Notice*

As noted above, humans can be contractually bound to website legal terms if there is inquiry notice. Notably, bots typically scrape every page of websites they visit unless disallowed by robots.txt (though, again, robots.txt compliance is merely voluntary at this moment). This means those bots likely scrape all pages with legal terms, including terms of service and terms of use. If a GenAI bot visiting the legal terms page of a website is not sufficient to make those terms binding under the same actual notice regime courts have consistently held for humans in contract law, then we are looking at yet another example of GenAI exceptionalism.

For instance, some may argue that though a bot not only visits the legal terms page but makes a complete copy of it, the terms should still not be binding on the bot's deployer because the bots do not understand the terms. However, such logic would imply that while *humans* do not need to understand terms they see or know about on a website for the terms to be legally enforceable, as is the case with almost all instances of clickwrap or enforceable browsewrap, bots are an extraordinary case and must understand the terms for the terms to be enforceable. In effect, we would be saying companies can subvert contract law by violating more terms and at a greater scale than by having an individual view a single legal terms page.

As explained above, an examination of whether website terms are enforceable usually looks at the conspicuousness of the terms, including the placement of the terms (are they before the "accept" or sign-in button), the conspicuousness of the links (underlines, different color, ALL CAPS, bold, etc.), whether it requires a click or two to accept, etc., rather than asking whether the user actually read the terms. However, the conspicuousness standard should be virtually eliminated when a bot scrapes the terms page. The least conspicuous links in the world would still be fully enforceable because the bot, and therefore the deployer, and therefore the GenAI company, has actual notice of the terms.

While this may seem extreme, it is a risk scrapers assume when they decide to crawl websites. It cannot be the case that if a human reads every page of a website the terms are binding, but the rule is easily subverted by having a bot visit the same pages on behalf of the human. If so, then no amount of conspicuousness matters for scraping.

For occasions when a bot does not visit the terms for whatever reason, we would then fall back to the inquiry notice standard. For bots with the ability to fill out forms and click buttons, any action on sign-in-wrap or click-wrap should be binding on the bot deployers for the same reason it is binding on a human: affirmatively agreeing to be bound by the terms. To believe otherwise is, again, to believe in GenAI exceptionalism. Any argument for GenAI exceptionalism must have some substantive policy justification.

## D. *Copyright Law*

If bots do not visit the legal terms, can they still be bound by browsewrap that meets even the most exacting and skeptical standards of a judge? Suppose a website has a large banner with prominent, high-contrast colors and bold font declaring that there are terms associated with the website, and that the link to the terms is at the bottom of the page. It seems highly doubtful that a reasonable human could argue they were unaware the site had legal terms. This is textbook inquiry notice. Therefore, a bot, which is a tool of the human deployers and thus merely performs the actions as directed by the humans, should bind the human to those terms as well.

But this introduces a new wrinkle. Many prominent copyright scholars, such as Mark Lemley, Oren Bracha, and Matthew Sag, have argued that AI companies should not be liable for copyright infringement in part because the software does not recognize or collect "expression" as they believe the terms should apply to copyright law. If bots don't recognize expression, they might argue, then even a bold, bright, unavoidable banner notifying visitors that there are ToS in ALL CAPS with underlined letters and flashing arrows pointing at it and though the banner follows the screen as the user scrolls, the bot would still not be under inquiry notice, and therefore no human or company could be bound by the terms for the bot's actions. Using a bot would obliterate, or at least entirely subvert, the commonsense understanding of providing notice to site visitors.

On the other hand, if common sense prevails and we apply contract law to bots the way we would apply it to humans, then bots can be liable for expression they encounter. It would seem odd to say a bot can be liable for expression under contract law but not copyright law. Therefore, enforcing contract law may simultaneously undermine a common key pillar of fair use arguments that software does not notice or care about expression, only the underlying facts and ideas.[65]

---

[65] This idea of an obvious and clear difference between how AI understands content and how humans understand it is, in the view of this author, unfounded. For instance, GenAI companies plainly prefer some training data over others even when the facts are the same. The reason? The expression of those facts is different. This implies, at a minimum, that GenAI picks up on the essence of expression, and could also imply that GenAI does, in fact, gain as much from the expression of the content as a human does.

## E. *Copyright Preemption*

Suppose a court finds the terms of a website binding on the bot deployers for whatever reason. Even so, those terms may not matter if the claim is based on lost revenue or something similar. The reason is copyright preemption.

With copyright preemption, copyright law's remedies trump the remedies of contract law when the right sought is "equivalent." Section 106(1) provides the most relevant language from the Copyright Act: "the owner of [a] copyright has the exclusive right[] to reproduce the copyrighted work," and Section 301(a): "no person is entitled to any such right or equivalent right in any such work under the common law or statutes of any State."

There is much dispute over what rights are "equivalent" to what the Copyright Act provides. Right now, under one analysis, it seems copyright never preempts contracts in the First, Fifth, Seventh, Eleventh, and Federal Circuits. Conversely, contracts are just presumptively not preempted in the Ninth Circuit. On the other hand, copyright presumptively preempts contracts in the Second Circuit. The remaining circuit foregoes a presumption and instead handles suits on a case-by-case basis.[66]

This means that if litigants bring a case before a Second Circuit judge claiming both copyright infringement and breach of contract, the judge will likely dismiss the contract claim. If the GenAI company goes on to win based on a fair use argument, then there is no infringement and no harm for site owners to seek relief.

For these reasons, if a website owner has concerns about a court finding the scraping of the site's content fair use despite restrictive or prohibitive ToS, we should expect the owner to try their hardest to have claims heard in the First, Fifth, Eleventh, Federal Circuit, and possibly the Ninth Circuit.

## VII. Social Implications: What It Means For Creators And Society

Despite what some GenAI boosters claim, a GenAI bot visiting a page is materially different from a human visiting. A bot could cause some very real damages that could become claims in a lawsuit. Moreover, any attempt to equate publicly available with public domain to justify the scraping or to minimize claims of damages is bound to run into trouble. This should be unsurprising. The internet economy was designed to accommodate humans, not bots.

There are several reasons a person may want their site to be easy to find and have few access barriers without implying that the site is fine with being scraped. For example, the person may want to generate revenue from the site, and a bot could interfere with or prevent the site from generating revenue.

---

[66]Kieran McCarthy, *Should Copyright Preemption Moot Anti-Scraping TOS Terms? (Guest Blog Post)*, TECHNOLOGY & MARKETING LAW BLOG (Dec 13, 2023), https://blog.ericgoldman.org/archives/2023/12/should-copyright-preemption-moot-anti-scraping-tos-terms-guest-blog-post.htm

GenAI bots themselves do not generate revenue for site owners. Scraping bots will not subscribe to a service or subscribe to access premium or other exclusive content, purchase merchandise, buy licensing rights, or enable the site to earn income by promoting other companies' products or services and earning a commission for each sale made through their website (*i.e.,* affiliate marketing).

Scraping bots also do a disservice to advertisers. If bots take content from a site and share a version of it through a GenAI platform, fewer people may visit the site itself, which means fewer people will see the advertisements on the site. This would mean fewer impressions and clicks on the advertisements, undermining the company whose products are being advertised and the advertising company displaying the ad.

The scraping bots also will not help a creator build a brand or reputation. Many people freely share content in the hopes of gaining the attention of potential partners or employers or creating a fan base or following. They may also share their content to establish their credibility as knowledgeable on a topic or with a particular set of skills. In any case, a bot scraping the content to serve on-demand in a similar but modified form, often without any attribution, undermines the site's purpose.

A site may also choose to be public so people can share information with others and build relationships with viewers or like-minded people. For example, the site may be for hobbyists to share their interests in a niche field and connect with others who share that passion, or they could share information to raise awareness of a topic.

| **Reasons Why a Person May Want Their Content to be Public While <u>*Not*</u> Wanting GenAI Bots to Scrape Them** ||
| --- | --- |
| **To Generate Revenue** | • Subscriptions <br> • Licenses <br> • Pay for premium content <br> • Order merchandise <br> • Pay for services (e.g., consultation) <br> • Affiliate marketing <br> • Ad revenue |
| **To Build a Brand or Reputation** | • To gain the attention of possible employers or an audience <br> • To gain the attention of possible partners <br> • To create a fan base or following <br> • To establish their credibility |
| **To Connect with Visitors** | • To build relationships with like-minded folks <br> • To share interests with fellow hobbyists |

|  | - To raise awareness of a topic |
|  | - To share an opinion and hope to convince others of a position |

Perhaps the only situation where we can presume scraping is ok, absent explicit consent, is for a site hosted by an anonymous person/group who publishes bespoke content anonymously, using a different pseudonym each time, and does not try to commercialize it in any way (ads, merchandise, subscriptions, etc.). But even then, the static domain name could be viewed as the site wanting to build a reputation as a place for people to share ideas anonymously without ulterior motives (like generating revenue).

Finally, for individual harms, suppose website owners feel they are merely unpaid cogs in the GenAI machine, producing desirable content that a bot will inevitably slurp up to enhance a model so an AI company can profit. In that case, people may feel disinclined to participate in the system, with bots chilling speech and other expression.

There could also be significant negative societal impacts. For any or all the reasons above, creators may feel that the only way to protect their work is by not making anything public (*e.g.*, by requiring an account or paywall). If they do limit access, it could impoverish society. Creators may have less incentive to create if they cannot profit. Less creation would mean there is less to share, and there would be a lower frequency of ideas exchanged. Cumulatively, this would undermine a range of Constitutional goals, such as promoting the progress of science and useful arts and fostering a robust and vibrant marketplace of ideas.[67]

This is not a hypothetical situation; it is the current reality. A paper by the Data Provenance Initiative examining the use of a dataset called C4, one of the most commonly used GenAI training datasets, states that "Our longitudinal analyses show that in a single year (2023-2024) there has been a rapid crescendo of data restrictions from web sources, rendering ~5%+ of all tokens in C4, or 28%+ of the most actively maintained, critical sources in C4, fully restricted from use. For Terms of Service crawling restrictions, a full 45% of C4 is now restricted."[68] It is difficult to imagine what could cause such a sudden and widespread clampdown on web scraping in 2023-2024 other than people rebelling against scraping for GenAI models specifically and not traditional scraping that has been ongoing for decades.

---

[67] Art 1, Sec 8, Cl 8; Abrams v. United States, 250 U.S. 616, 630 (1919), Justice Holmes dissenting ("…the ultimate good desired is better reached by free trade in ideas -- that the best test of truth is the power of the thought to get itself accepted in the competition of the market, and that truth is the only ground upon which their wishes safely can be carried out. That, at any rate, is the theory of our Constitution.")

[68] Shayne Longpre et al., *Consent in Crisis: The Rapid Decline of the AI Data Commons*, https://www.dataprovenance.org/Consent_in_Crisis.pdf

How courts interpret other legal concepts, like contract preemption, could also have an outsized effect on creators and the advancement of science and art. If courts neuter contract claims by having copyright preempt them, and virtually all GenAI use of copyrights is deemed fair use, then the arrangement would seem to incentivize people to hide more content from the public (e.g., putting it behind a log-in), which would make it more difficult to find via web searchers. It would essentially mean that you consent to being scraped by anyone for GenAI purposes unless you actively hide yourself. The default mode of the internet could become closed communities rather than open ones. This would make generating meaningful revenue from subscriptions, ads, selling stuff, and so on challenging or impossible, disrupting commercial enterprises and innovation.

## VIII. A Scraping Exception

Many sites have become so flummoxed by scraping bots that they now disallow *all* bots via robots.txt. Reddit is one example. Its content has proven highly valuable for GenAI training, so it now disallows all bots but is willing to make licensing deals with companies like Google to access the content.[69] Reddit also notes that it will view nonprofits favorably, giving them access to content for free. However, not every site has Reddit's resources to disentangle various bots. As a result, when most sites disallow all bots, they make no distinction between nonprofits, archives, and academic researchers on the one hand and for-profit GenAI companies on the other.

Even when sites try to be more nuanced in their terms by specifying that the content can be scraped for non-commercial use, they tend to have more restrictive robots.txt because robots.txt is incapable of communicating the same nuance, meaning the bots respecting robots.txt can't access the content even if they were going to comply with the ToS.[70]

---

[69] Emma Roth, *Reddit is now blocking major search engines and AI bots — except the ones that pay*, THE VERGE (Jul 24, 2024), https://www.theverge.com/2024/7/24/24205244/reddit-blocking-search-engine-crawlers-ai-bot-google

[70] Shayne Longpre et al., *Consent in Crisis: The Rapid Decline of the AI Data Commons*, https://www.dataprovenance.org/Consent_in_Crisis.pdf

There are several reasonable arguments for why nonprofit research organizations should be exempt from some laws, including contract law for web scraping of public data. For instance, the scraped data is necessary for their work on projects that help advance public knowledge and address social issues. These social issues often concern valuable areas such as public health, education, and social justice. Typically, nonprofit research organizations are bound by ethical guidelines that help ensure data is used responsibly while fostering innovation. They are also more transparent, publishing their work and the materials their work is based on, enabling greater accountability. Finally, with some exemptions, nonprofits will likely be able to perform the necessary research because they often have limited resources (talent, funding, legal representation, etc.).

Given that data is the lifeblood of GenAI and any research of highly-capable GenAI will require access to data traditionally gathered by sites like Common Crawl, it may make sense to create a blanket exemption or some other allowance for bots deployed exclusively to promote nonprofit scientific research under certain conditions. An exemption should probably require at least three criteria: (i) the data must be released under a non-commercial license, (ii) the data must be released under a license that allows for scientific use and research only, and (iii) the data must gate access to the data and models (verify identities of people who want to download the models/datasets), such as with ROOTS and the BigCode PII training dataset. Otherwise, large, high-quality datasets will almost certainly enable and implicitly encourage data laundering.

## IX. Recommendations And Conclusion

Given the dramatic shift in the aggressive nature of scraping, the voracious appetite of companies (especially GenAI companies) for data, and the potential for far-reaching and foreseeable harms to content owners, it is time to rethink robots.txt and contract law on the internet.

As the Data Provenance Initiative notes:

> [Robots.txt] places an immense burden on website owners to correctly anticipate all agents who may crawl their domain for undesired downstream use cases. We consistently find this leads to protocol implementations that don't reflect intended preferences. An alternative scheme might give website owners control over *how* their webpages are used rather than *who* can use them. This would involve standardizing a taxonomy that better represents downstream use cases, *e.g.* allowing domain owners to specify that web crawling only be used for search engines, or only for non-commercial AI, or only for AI that attributes outputs to their source data. New commands could also set extended restriction periods given dynamic sites may want to block crawlers for extended periods of time, *e.g.* for journalists to protect their data freshness. Ultimately, a new protocol should lead to website owners having greater capacity to self-sort consensual from non-consensual uses, implementing machine-readable instructions that approximate the natural language instructions in their Terms of Service.[71]

Whatever the outcome of any changes to robots.txt or its successors, contract law must remain the determining factor in any dispute about how others may use website content. Robots.txt is a handy shorthand for a website to indicate its preferences, and it is a simple and cost-efficient way to direct bots, but it is not a sufficient replacement for contract law. Until companies can convincingly argue that their scraping of data overwhelmingly benefits society more than it harms society, we should continue to favor human autonomy and control of the content humans create rather than seek to make life as simple and frictionless for for-profit companies as possible. Companies are not entitled to easy money and free access to raw materials.

On the other hand, we must retain looser restrictions on nonprofit organizations that primarily focus on scientific research and share their data and other research responsibly to minimize the negative impact on content creators. Without their efforts and reasonably frictionless access to data on which to conduct research, society risks conducting too little research and obtaining too little insight into profoundly important areas of science and art.

In sum, a general position of granting GenAI bots greater leeway under the law than humans is a terrible policy. It makes it too easy to subvert the purpose of contract law, unfairly disrupts or destroys markets, introduces unnecessary confusion, and adds unnecessary and stifling burdens on the exchange of ideas.

---

[71] Shayne Longpre et al., *Consent in Crisis: The Rapid Decline of the AI Data Commons*, https://www.dataprovenance.org/Consent_in_Crisis.pdf